\documentclass[graybox, envcountchap]{svmult}
\usepackage{tikz}
\usetikzlibrary{arrows.meta,patterns}

\usepackage{mathptmx}        
\usepackage{amsmath, bbm}
\usepackage{amssymb}
\usepackage{color}
\usepackage{helvet}          
\usepackage{courier}         
\usepackage{dirtree}
\usepackage{makeidx}        
\usepackage{graphicx}        
\usepackage{subfig}

\usepackage{multicol}        
\usepackage[bottom]{footmisc}

\usepackage{pgfplots}
\pgfplotsset{compat=1.18}

\usepackage{hyperref}        
\hypersetup{colorlinks=true,urlcolor=blue}

\usepackage[misc]{ifsym}

\makeindex             

\newcommand{\id}{\mathbbm{1}}

\newcommand{\ket}[1]{\left | \, #1 \right \rangle}
\newcommand{\bra}[1]{\left \langle #1 \, \right |}

\newcommand{\tr}{\mathrm{tr}}

\newcommand{\be}{\begin{equation}}
\newcommand{\ee}{\end{equation}}
\newcommand{\ba}{\begin{eqnarray}}
\newcommand{\ea}{\end{eqnarray}}
\usepackage{url}

\begin{document}

\title{Page Curve of average subsystem entropy }
\author{Oscar C. O. Dahlsten}
\institute{City University of Hong Kong, 88 Tat Chee Avenue, Kowloon Tong, Hong Kong SAR, China \email{oscar.dahlsten@cityu.edu.hk}}

\maketitle

\abstract{The Page curve is a curve of average subsystem entropy as a function of subsystem size. The curve starts from 0, rises, peaks near the maximal possible entropy when the subsystem makes up half of the total system, and then falls back to 0. We here describe subsystem entropy, averaging over quantum states, and why the curve rises and falls in that manner. We also discuss the connection between the curve and the black hole information paradox.}


\vspace{2pc}

\section{Introduction}
\begin{figure}
\begin{tikzpicture}
\begin{axis}[
    width=12cm,
    height=8cm,
    xlabel={Logarithm of Subsystem Dimension},
    ylabel={Subsystem Entropy},
    title={Page Curve},
    grid=major,
    legend pos=north west,
    xmin=0,
    xmax=100,
    ymin=0,
    ymax=100
]

\addplot[
    domain=0:100,
    blue,
    thick
] {x};

\addplot[
    domain=0:50,
    red,
    thick,
    dashed
] {x};
\addplot[
    domain=50:100,
    red,
    thick,
    dashed
] {100-x};

\legend{Semiclassical Curve, Page Curve}

\end{axis}
\end{tikzpicture}
\caption{{\bf Page curve.} The entropy of a subsystem, averaged over the uniform distribution over pure quantum states, rises with the subsystem size, peaks when the subsystem is half of the total subsystem (50/100 subsystems in this case) and then falls off in a symmetric manner down to 0. The Page Curve contrasts with the behaviour of the entropy curve of black hole radiation derived semiclassically by Hawking and others. The latter curve increases monotonically as the radiation increases, suggesting a black hole evaporation leaves behind high entropy radiation rather than a pure state radiation field.}
\label{fig:Pagecurve}
\end{figure}
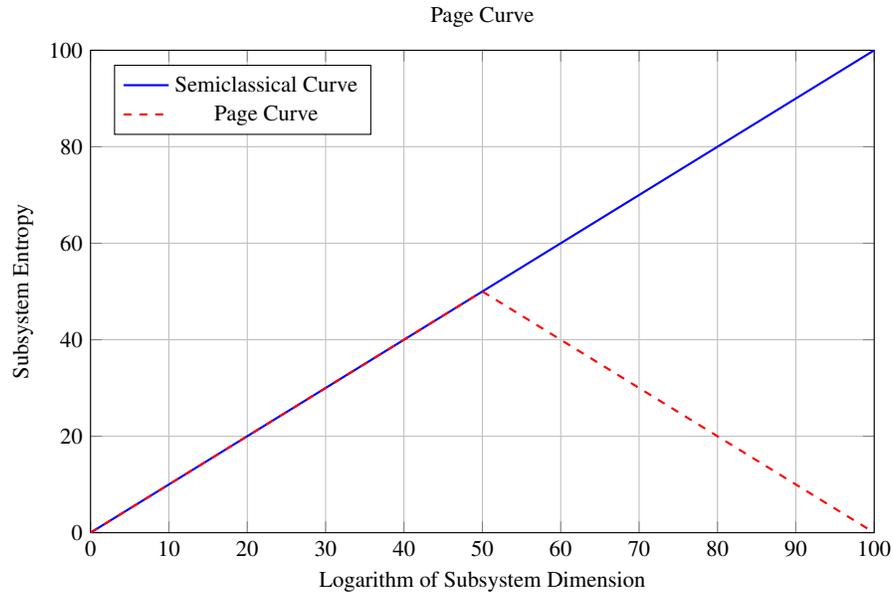

The Page curve concerns the expected value of subsystem entropy. More specifically, for a bipartite pure state $\ket{\Psi}_{AB}$, there is an entropy of subsystem $A$ for example, $S(\rho_A)$ where $\rho_A$ is the reduced state on $A$. The Page curve shows the average $S(\rho_A)$, averaged over pure bipartite states, as a function of $\log d_A$ where $d_A$ is the dimension of system $A$. The curve is depicted in Fig.\ref{fig:Pagecurve}. The curve likely gained its name because of the 1993 papers Refs.\cite{page1993average,page1993information}. Similar results concerning the curve were obtained by Lubkin around 1978 in Ref.~\cite{lubkin1978entropy} and consistent results were derived in Ref.~\cite{lloyd1988complexity} by Lloyd and Pagels around 1988. Page's early work appears to have been motivated by wishing to show that entropy does not necessarily change monotonically as subsystems radiate out as suggested by so-called semi-classical Hawking radiation calculations. The curve and related concepts also appear in quantum thermodynamics and quantum information science.  

This chapter aims to give an introduction to the curve. The reader is assumed to have graduate knowledge of quantum theory. The emphasis will be on the quantum information science side, reflecting the expertise of the author. The chapter is far from an exhaustive review of the literature. (For a more developed review on the expected entropy of subsystems see e.g. Ref.~\cite{dahlsten2014entanglement})

We proceed as follows. We begin with entropy and subsystem entropy. We then describe averaging over states, before showing an example of how one can derive the curve. Finally we give a discussion on the relation to black holes, thermodynamics and quantum information science. 

\section{Subsystem entropy}
In this section, we first discuss the entropy of quantum states, including Renyi entropy, and then the entropy of subsystems. 
\subsection{Entropy of quantum states}
The entropy of a quantum state $\rho$ is a single real number assigned to the state.  An important family of possible entropy choices are the quantum Renyi entropies 
\begin{equation}
S_q(\rho)=\frac{1}{1-q}\log \sum_{i=1}^d \lambda_i^q,
\label{eq:Sq}
\end{equation}
where $\lambda_i$ denotes the i-th eigenvalue of the density matrix $\rho$, i.e.\ $\rho=\sum_{i=1}^d\lambda_i\ket{i}\bra{i}$. The most commonly employed entropy
\begin{equation}
S_{q\rightarrow 1}(\rho):=S_1(\rho):=S(\rho)=\sum_{i=1}^d\lambda_i \log \frac{1}{\lambda_i},
\label{eq:S1}
\end{equation}
is often termed the {\em von Neumann entropy} after a thermodynamical argument leading von Neumann to this definition (see Ref.~\cite{peres1997quantum} for the argument). In information theory the logarithm $\log(.)$ is normally taken to base 2 and in thermodynamics to base e=2.72....Furthermore in thermodynamics the entropy is by definition also multiplied by Boltzmann's constant $k_B$, becoming $k_B\sum_{i=1}^d\lambda_i \ln \frac{1}{\lambda_i}$. For our purposes here, it will also be useful to note that, from Eq.~\ref{eq:S1}, 
\begin{equation}
S_2=-\log \sum_{i=1}^d \lambda_i^2,
\label{eq:S2}
\end{equation}
where $\sum_i \lambda_i^2\in [\frac{1}{d},1] $ is commonly termed the {\em purity} of the state $\rho$. 
 
The entropy can be understood in multiple consistent manners: 
\begin{itemize}
\item  $S(\rho)\in [0,\log d]$ quantifies our lack of information about the preparation of the quantum system. $S(\rho)=0$ if and only if the state has form $\rho=\ket{i}\bra{i}$ for some state $\ket{i}$ which then has eigenvalue 1. $S(\rho)=\log d$ for $\rho=\sum_{i=1}^d\frac{1}{d}\ket{i}\bra{i}$ which corresponds to no knowledge of the preparation, the maximally mixed state. $S(\rho)$ can be interpreted as quantifying where on the scale between these two extremes the state  $\rho$ lies.  
\item  $S(\rho)$ is the classical entropy $H(\vec{\lambda})=\sum_i p_i \log\frac{1}{p_i}$ of a measurement in the eigenbasis of $\rho$ whereby $\lambda_i=\tr(\rho\ket{i}\bra{i})=:p_i$ for $\rho=\sum_i \lambda _i\ket{i}\bra{i}$.
\item $S(\rho)$ is the classical entropy $H$ of measurement in a given basis {\em minimised over all bases}: 
\begin{equation}
S(\rho)=\min_{\{\ket{j}\bra{j}\}_{j=1}^d}H(\vec{p}), 
\end{equation}
where $\vec{p}$ is the vector of probabilities with entries $p_j=\tr(\rho \ket{j}\bra{j})$.
\end{itemize}
For a tutorial with more on entropies see e.g.\ Ref.~\cite{dahlsten2025entropy}.

\subsection{Entropy of a subsystem}
The entropy of a subsystem A of a joint system AB is given by $S(\rho_A)=S(\tr_B\rho_{AB})$ where $\tr_B$ is the partial trace over B, the quantum method of taking a marginal distribution on A. $S(\rho_A)\in [0,\log d_A]$ where $d_A$ is the state dimension of $A$. $\log d_A$ can be interpreted as the number of qubits, $n_A$, making up $A$. $n_A$ qubits of dimension 2 have dimension $d_A=2^{n_A}$ such that $n_a=\log d_A$.

The subsystem entropy of a pure state can be understood via the singular value decomposition of a matrix representing the state. The singular value decomposition of a complex matrix $M$ is that $M=UDV$ where $U$ and $V$ are unitary and $D$ is diagonal with non-negative real entries. At the component level $M_{ij}=\sum_k U_{ik}D_{kk}V_{kj}:=\sum_kU_{ik}\mu_{k}V_{kj}$. Thus, a pure state on AB can be rewritten as
\begin{equation}
\ket{\Psi}_{AB}=\sum_{i,j,k}M_{ij}\ket{i}_A\ket{j}_B=\sum_{i,j}U_{ik}\mu_{k}V_{kj}\ket{i}\ket{j}=\sum_{i,j,k}\mu_{k}(U_{ik}\ket{i})(V_{kj}\ket{j}):=\sum_{k}\mu_{k}\ket{\tilde{k}}_A\ket{\tilde{k}}_B,
\end{equation}
which is termed the {\em Schmidt decomposition} of $\ket{\Psi}_{AB}$ with Schmidt coefficients $\mu_{k}$. 
It follows that
\begin{equation}
\rho_A= \sum_{k}\mu_{k}^2 \ket{\tilde{k}}_A\bra{\tilde{k}}_A
\end{equation}
and 
\begin{equation}
\rho_B= \sum_{k}\mu_{k}^2 \ket{\tilde{k}}_B\bra{\tilde{k}}_B.
\end{equation}
Thus A and B share the same eigenvalues, $\mu_{k}^2$ and accordingly $S(\rho_A)=S(\rho_B)$,  for this case of a pure bipartite joint state. 

$S(\rho_A)$ is, in the case of a pure bipartite state, the standard measure of the entanglement between $A$ and $B$. A quick way to understand why is as follows. By definition, entangled states are those which are not product states (or probabilistic mixtures thereof). Pure product states have Schmidt form $\ket{1}\ket{1}$. On the other extreme, the maximally entangled state is by definition  $\sum_{k=1}^{d_A}\frac{1}{\sqrt{d_A}}\ket{k}\ket{k}$. $S(\rho_A)\in [0,\log d_A]$, can be interpreted as quantifying where on the scale between these two extremes the state on AB is.  $S(\rho_A)=0$ for product states and $S(\rho_A)=\log d_A$ for the maximally entangled state. (For mixed states, entanglement is quantified by other measures that distinguish between quantum and classical correlations--see e.g. Ref.\cite{vedral1997quantifying}).

\section{Averaging over quantum states}
In this section, we first argue that the uniform distribution for fair averaging over states should be unitarily invariant. We then say more about what that unitarily invariant distribution is.

\subsection{Uniform distribution should be unitarily invariant}
For fair averaging over quantum states we require a uniform distribution over quantum states. A uniform distribution can be defined via invariance under some permutation or more general group operation. For example for a coin with heads H or tails T, [p(H), p(T)] is uniform if and only if it is invariant under a permutation of p(H) and p(T). Now instead of H and T let the microstates be pure quantum states. There is a continuum of these so we are looking for a uniform  probability density over those states. 

To visualise distributions over pure quantum states consider the 3-dimensional real space Bloch sphere representation of a qubit, where $\rho$ is represented via the expectation value vector $(\langle X\rangle, \langle Y\rangle, \langle Z\rangle)^T$  (T is the tranpose). Recall that the Bloch vector uniquely determines the normalised density matrix since the Paulis plus the identity, $\{\id, X,Y,Z\}$, form a basis for Hermitian matrices such that 
\begin{eqnarray}
\rho &=&\xi_0\id+\xi_1X+\xi_2Y+\xi_3Z;\\  
\tr(\rho\id)&=& 2\xi_0=1;\, \langle X\rangle=\tr(\rho X)=2\xi_1 \,\,\textrm{etc.}.
\end{eqnarray}
Valid qubit states respect 
\begin{equation}
\label{eq:sphere}
\langle X\rangle^2+ \langle Y\rangle^2+ \langle Z\rangle^2\leq 1,
\end{equation}
with equality for pure states which thus lie on the surface of a sphere. It turns out that each point on the sphere surface (for a single qubit) is a valid state. $\langle X\rangle^2\in [0,1]$ can be interpreted as the predictability of the observable $X$ which has eigenvalues $\pm 1$, being 0 for $p(1)=p(-1)$. Thus Eq.~\ref{eq:sphere} is sometimes interpreted as an uncertainty relation, bounding the joint predictability of the three Paulis~\cite{brukner2009information}.

\begin{figure}

\begin{tikzpicture}
    \begin{scope}
        \clip (-4-2,0-2) rectangle (-4+2,0+2);
        \fill[gray!20] (-4,0) circle (2);
        \fill[pattern=north west lines, pattern color=black] (-4,0.5) ellipse (1.8 and 0.8);
        \fill[pattern=north west lines, pattern color=black] (-4,1.2) ellipse (1.4 and 0.5);
        \draw (-4,0) circle (2);
        \draw[thick] (-4,0) circle (0.2);  
        \fill (-4,0) circle (0.08);        
        \node[above right] at (-4,0) {$\langle Z \rangle$};
    \end{scope}

    \begin{scope}
        \clip (4-2,0-2) rectangle (4+2,0+2);
        \fill[gray!20] (4,0) circle (2);
        \begin{scope}[rotate around={45:(4,0)}]
            \fill[pattern=north west lines, pattern color=black] (4,0.5) ellipse (1.8 and 0.8);
            \fill[pattern=north west lines, pattern color=black] (4,1.2) ellipse (1.4 and 0.5);
        \end{scope}
        \draw (4,0) circle (2);
        \draw[thick] (4,0) circle (0.2);   
        \fill (4,0) circle (0.08);         
        \node[above right] at (4,0) {$\langle Z \rangle$};
    \end{scope}

    \draw[-{Stealth[length=3mm,width=2mm]}, thick] 
        (-1.5,0) -- node[above] {$U$} (1.5,0);

\end{tikzpicture}
\caption{{\bf Non-uniform density on Bloch sphere.} Toy picture of how non-uniform density of pure states changes under unitary evolution. The circle represents the Bloch sphere real vector representation of a qubit as seen from above the z-axis. The shading indicates density. The unitary acts as a rotation, altering the overall density. In the case depicted, the density changes under rotation and is thus not unitarily invariant. In contrast, a uniform density would be rotationally invariant.}
\label{fig:nonuniform}
\end{figure}
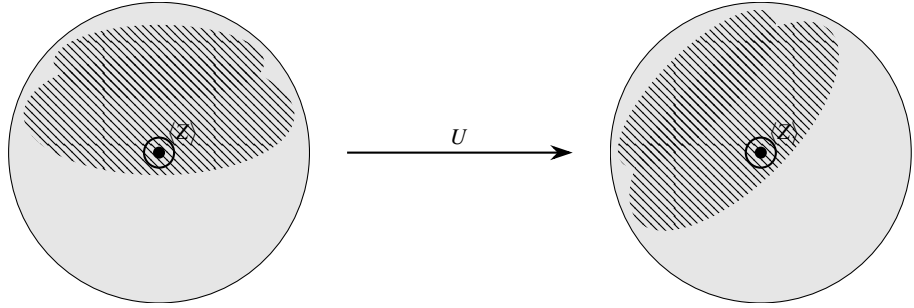

The analogue of a microstate permutation is now a a rotation of the sphere. That rotation is equivalent to a unitary evolution $U$ being applied on the standard complex vector representation of the state. The analogue of a uniform distribution is then a distribution which is invariant under any unitary: {\em the unitarily invariant distribution}. This distribution can be visualised as a uniform density over the Bloch sphere surface. If the distribution were not uniform, rotations of the sphere would alter it, as depicted in Fig.~\ref{fig:nonuniform}. 

The above discussion motivates using unitary invariance to define a uniform distribution (called measure) over quantum states. More concretely, for whatever parameterisation $(\theta_1,\theta_2,...)$ is used for the state, the joint density over those parameters $\rho (\theta_1,\theta_2,...):=\rho(\ket{\phi})$ should respect
\begin{equation}
\rho(\ket{\phi})=\rho(U\ket{\phi}) \,\,\, \forall U\,|\,U^{\dagger}U=\id.
\end{equation}

\subsection{Unitarily invariant distribution over pure states}
To gain intuition about how to pick states randomly uniformly, consider firstly picking real vectors in a manner such that the density is invariant under orthogonal matrices. In particular, consider again the case of a single qubit and the Bloch sphere, wherein states are real-vectors and unitaries act as orthogonal matrices. Let for notational simplicity $x:=\langle X\rangle$ etc..  We want a method for picking the real vectors $(x,y,z)^T$ with probability density respecting $\rho(x,y,z)=\rho(x',y',z')$ if $(x',y',z')^T=\mathcal{O}(x,y,z)^T$ where $\mathcal{O}$ is an orthogonal matrix ($\mathcal{O}^T\mathcal{O}=\id$). Thus $\rho(x,y,z)$ should at most depend on the radius $r=(x^2+y^2+z^2)^{1/2}$ since otherwise we could alter the density via some $\mathcal{O}$ that keeps $r$ constant but alters the angles. A simple way to have $\rho(x,y,z)$ only depending on the radius is to pick $x,y,z$ independently from the normal distribution such that $\rho(x,y,z)=\mathcal{N} \exp(-x^2)\exp(-y^2)\exp(-z^2)=\mathcal{N}\exp(-r^2)$, where $\mathcal{N}$ is the normalisation of the joint probability distribution. This is then an orthogonally invariant density (often called measure). A subtlety is that for pure quantum states we also require r=1. We then add a second step to alter $r$, by a deterministic many-to-one map $x\rightarrow r^{-1}x$ etc.. The density of the altered values is by inspection still orthogonally invariant since it is independent of the angles. 

Now consider the most common parametrisation of the pure quantum state as a complex vector $\ket{\phi}$ with entries $\phi_j=a_j+ib_j$. It turns out that picking the state from the unitarily invariant distribution means picking all the real components $a_j$ and $b_j$ for all $j$ independently from the normal distribution and finally normalising the vector~\cite{diaconis2005random}, similar to the above simple case of real numbers.  

Consider further the distribution over individual observable expectation values $\langle g_i\rangle$. If the state is picked at random, $\langle g_i\rangle$, or any function thereof such as $\langle g_i\rangle^2$ is picked at random. Recall that $\langle g_i\rangle^2\in [0,1]$ can be interpreted as the predictability of $g_i$ for $g_i$ having eigenvalues $\pm 1$. There is an associated `expectation value of the expectation value', such as $\mathbb{E}(\langle g_i\rangle^2)$. These values, being functions of the density over states, should be invariant for observables connected by unitaries e.g.\ $\mathbb{E}(\langle X\rangle^2)=\mathbb{E}(\langle Z\rangle^2)$. If that equality were not the case, we could alter the density over state parameters $\langle X\rangle$ and $\langle Z\rangle$ by applying the unitary that connects the two observables ($Z=HXH$ where $H$ is the Hadamard unitary). This equality of expected predictability across unitarily related observables is helpful in calculations and in understanding why expected subsystem entropy is high.

\section{Average subsystem entropy}
We are now ready to discuss the average subsystem entropy. We give an example of one way to  derive the expected purity of a subsystem. We argue that the total predictability is limited and that the expected share thereof which is associated with the local observables is small, such that the local purity is small.

We shall focus on the predictability of Pauli matrix observables (generalised to $n$ qubits). This is an approach that the author and collaborators have found convenient for understanding and for undertaking calculations. We used a similar approach to show that states generated by random collections of 2-qubit unitaries of restricted (poly($n$)) length generate approximately the same average subsystem entropy~\cite{oliveira2007generic} as those picked from the unitarily invariant distribution. Furthermore, we used a similar approach to show that there is a classical analogue of states typically being maximally entangled (more details below).

\subsection{Total predictability limited}
The predictability of observables is limited~\cite{brukner2009information}. Consider the implication for states written in a Pauli basis for Hermitian matrices $\rho_{AB}=\sum_i\xi_ig_i$ where $g_i\in \{X,Y,Z,\id\}^{\otimes n}$. We can, similarly to the above single qubit case, interpret $\langle g_i\rangle^2\in [0,1]$ as the predictability of $g_i$. The purity determines the total predictability in that 
\begin{equation}
\tr(\rho^2) = d \sum_{i=0}^{d^2-1} \xi_i^2= \frac{1}{d}\sum_{i=0}^{d^2-1} \langle g_i\rangle^2=\frac{1}{d}(1+\sum_{i=1}^{d^2-1}\langle g_i\rangle^2),
\label{eq:puritybounded}
\end{equation}
where $d=2^n$ and we used 
${\rm Tr} (g_j g_k) = d \delta_{jk} \quad \forall j,k.$. Furthermore, the first element $g_0=I_{2^n}$ has a fixed coefficient $\xi_0=1/d$ since $\tr{\rho}=1$. 

Eq:\ref{eq:puritybounded} by inspection implies, for pure states where $\tr(\rho^2)=1$,
\begin{equation}
\sum_{i=1}^{d^2-1} \langle g_i\rangle^2= d-1,
\label{eq:predictabilitybounded}
\end{equation}
which is a strong limitation since the algebraic maximum of the left-hand-side is $d^2-1$. Thus the total predictability is strongly limited. 

\subsection{Expected local predictability a small share of total}

There is a system $AB$ comprising a number $n=n_A+n_B$ of individual qubit systems. We can break Eq.~\ref{eq:predictabilitybounded} into one contribution from local observables on A and another from the rest.  Local observables on A (B) correspond to measuring A (B)  alone and take form $g_A\otimes \id_B$ ($\id_A\otimes g_B$). Thus,
\begin{equation}
\sum_{g_i|g_i=g_i^{(A)}\otimes \id_B} \langle g_i\rangle^2+\sum_{g_i|g_i\neq g_i^{(A)}\otimes \id_B} \langle g_i\rangle^2= d-1,
\label{eq:purityboundedlocal}
\end{equation}
The first term associated with local observables on A can be termed the local predictability on A and equals $\sum_i\langle g_i^{(A)}\rangle^2$ since $\langle g_i^{(A)}\rangle=\langle g_i^{(A)}\otimes \id_B \rangle$. 


There are many more global observables than local. Global Observables are not of that local form, e.g. $X_A\otimes Z_B$ can be called global. When we use the Pauli bases for systems composed of many qubits $\{X,Y, Z, \id \}^{\otimes n}$, we have four choices for each qubit so $4^n$ basis elements. There are similarly $4^{n_A}$ basis elements local on $A$ so the ratio of local on A to global is $4^{n_A-n}$. This ratio decreases exponentially in $n$ for all $n_A$, since for the largest A of interest, $n_A=n/2$ the ratio becomes $4^{-n/2}$. 

Consider the expected share of predictability split in a similar manner. As argued above, for unitarily invariant statistics 
\begin{equation}
\mathbb{E}(\langle g_i\rangle^2)=\mathbb{E}(\langle g_j\rangle^2):=\mathbb{E}(\langle g\rangle^2)\,\,\,\mathrm{for}\, i,j\neq 0.
\label{eq:giequal}
\end{equation}

Eq.\ref{eq:predictabilitybounded} and Eq.\ref{eq:giequal} together give
\begin{equation}
(d^2-1)\mathbb{E}(\langle g\rangle^2)= d-1,
\end{equation}
such that 
\begin{equation}
\mathbb{E}(\langle g\rangle^2) = \frac{d-1}{d^2-1}.
\end{equation}

Thus the expected predictability on A,
\begin{equation}
\mathbb{E}\left(\sum_{g_i|g_i=g_i^{(A)}\otimes \id_B} \langle g_i\rangle^2\right)=(d_A^2-1)\mathbb{E}(\langle g\rangle^2)= \frac{(d_A^2-1)(d-1)}{d^2-1},
\label{eq:predictabilityAbounded}
\end{equation}
which is small compared to the total predictability of $(d^2-1)\mathbb{E}(\langle g\rangle^2)=d-1$.

\subsection{Expected local purity low, entropy high}


The expected purity on A is closely related to the expected predictability of Eq.~\ref{eq:predictabilityAbounded}. From eq.\ref{eq:puritybounded}
\begin{equation}
\mathbb{E}\left(\tr(\rho_A^2)\right)=\frac{1}{d_A}\left(1+\mathbb{E}\left(\sum_{i\neq 0}\langle g_i\rangle^2\right)\right)= \frac{1}{d}\left(1+\frac{(d_A^2-1)(d-1)}{d^2-1}\right)
\label{eq:expectedApurity}
\end{equation}


Eq.~(\ref{eq:expectedApurity}) indeed simplifies to Lubkin's expression (derived differently)~\cite{lubkin1978entropy}:
\begin{equation}
\mathbb{E}\left(\tr(\rho_A^2)\right)= \frac{d_A+d_B}{d_Ad_B+1},
\label{eq:LubkinPurity}
\end{equation}
where total system dimension $d=d_Ad_B$. We see that $\mathbb{E}\left(\tr(\rho_A^2)\right)=1$ (maximal purity) when $d_A=d$ or $d_B=d$ and moreover, by differentiation one sees  $\mathbb{E}\left(\tr(\rho_A^2)\right)$ has a minimal value, for fixed $d$ at $d_A=\sqrt{d}$. Thus the entropy, which essentially has an inverse behaviour to the purity has the qualitative behaviour of rising and falling in Fig~\ref{fig:Pagecurve}. A quick way to relate purity $\tr(\rho^2)$ to entropy is to note that $S_2=-\log \tr(\rho^2)$. (A subtlety is that there is a spread around the average but that spread can be shown to be heavily limited~\cite{lubkin1978entropy}.) 

Similar arguments can be made for the expected subsystem purity/entropy of classical systems~\cite{muller2012unifying}.  For the case of classical probability distributions on bits we would drop X and Y so $\rho$ is diagonal and there are $2^n$ global observables and $2^{n_A}$ on $A$. The total predictability, which may be limited, is again shared out between observables and similar arguments apply~\cite{muller2012unifying}. One may say that, whether in quantum or classical systems, when there is limited predictability, it tends to be found in global observables, i.e. in correlations, rather than local observables.

\section{Discussion}

\subsection{Expected subsystem entropy and black holes}\label{thermosec}

Page's seminal work \cite{page1993information} applied expected subsystem entropy ideas to black hole thermodynamics. Page modelled formation and evaporation as a unitary process. The entropy of the radiation then typically increases until what is now termed the Page time (when the dimensions of the inside and outside are equal), then decreases to zero. This model has information preservation, due to the reversibility of unitary evolution, while appearing random/thermal locally. This resolves the apparent paradox where calculations showing entropy increase in reduced states of subsystems might suggest information loss, by noting that information can be preserved in correlations even when subsystems appear maximally mixed. 

If the essential picture of black holes radiating away is correct and the radiation entropy follows something like the Page curve, the generalised 2nd law hypothesized by Hawking~\cite{hawking1975particle}  that $S+\frac{A}{4}$ never decreases where $S$ is the entropy of matter outside the black holes and $A$ is the sum of the surface areas of the event horizons would, at least taken on face value, be violated, as $A$ goes to 0 during evaporation and $S$ goes to 0 at the end of the Page curve. It should finally be noted that Page~\cite{page1993information} acknowledges that there is a key question mark around how the subsystems could exit the black hole to facilitate this curve, a question that remains an active area of research.  

\subsection{Expected subsystem entropy as an approach to thermodynamics}\label{thermosec}

As noted by Feynman ``{\em This fundamental law [that systems are in a Gibbs thermal state] is the summit of statistical mechanics, and the entire subject is either the slide-down from this summit, as the principle is applied to various cases, or the climb-up to where the fundamental law is derived and the concepts of thermal equilibrium and temperature $T$ clarified}''~\cite{feynman2018statistical}.

The mathematical fact that most pure bipartite quantum states are nearly maximally entangled provides an alternative approach to justify the thermal state assumption \cite{lloyd1988complexity, lubkin1978entropy, gemmer2001quantum, popescu2006entanglement}. For the simplest case where $H=0$, the Gibbs state reduces to $\rho_{th}(\beta, H)=\frac{\id_{d_A}}{d_A}$. This matches the reduced state ${\tr}_B \rho_{AB}$ when $\rho_{AB}$ is pure and maximally entangled. The typically maximal entanglement then implies that the system should typically be in a thermal state.

For non-trivial Hamiltonians, the argument is more subtle. One may pick a state from a uniform distribution over a restricted state space wherein the total energy on AB is restricted to a particular value. Then local observables are still not expected to have predictability such that the local reduced state is the same as though we had taken a maximally mixed state in the restricted state space, as in the more common derivation of the Gibbs state (see e.g.\ Ref.~\cite{dahlsten2014entanglement} for references). Such arguments give an alternative justification for the Gibbs state.

\bibliographystyle{plain} 
\bibliography{refs} 

\begin{thebibliography}{10}

\bibitem{brukner2009information}
{\v{C}}aslav Brukner and Anton Zeilinger.
\newblock Information invariance and quantum probabilities.
\newblock {\em Foundations of Physics}, 39:677--689, 2009.

\bibitem{dahlsten2025entropy}
Oscar Dahlsten.
\newblock Tutorial 1: Entropy and majorization in generalized probabilistic theories.
\newblock \url{https://youtu.be/fXGUiXur1kU?feature=shared}, Accessed Feb 2025.
\newblock YouTube Video.

\bibitem{dahlsten2014entanglement}
Oscar~CO Dahlsten, Cosmo Lupo, Stefano Mancini, and Alessio Serafini.
\newblock Entanglement typicality.
\newblock {\em Journal of Physics A: Mathematical and Theoretical}, 47(36):363001, 2014.

\bibitem{diaconis2005random}
Persi Diaconis.
\newblock What is a random matrix.
\newblock {\em Notices of the AMS}, 52(11):1348--1349, 2005.

\bibitem{feynman2018statistical}
Richard~P Feynman.
\newblock {\em Statistical mechanics: a set of lectures}.
\newblock CRC press, 2018.

\bibitem{gemmer2001quantum}
Jochen Gemmer, Alexander Otte, and G{\"u}nter Mahler.
\newblock Quantum approach to a derivation of the second law of thermodynamics.
\newblock {\em Physical Review Letters}, 86(10):1927, 2001.

\bibitem{hawking1975particle}
Stephen~W Hawking.
\newblock Particle creation by black holes.
\newblock {\em Communications in mathematical physics}, 43(3):199--220, 1975.

\bibitem{lloyd1988complexity}
Seth Lloyd and Heinz Pagels.
\newblock Complexity as thermodynamic depth.
\newblock {\em Annals of Physics}, 188(1):186--213, 1988.

\bibitem{lubkin1978entropy}
Elihu Lubkin.
\newblock Entropy of an n-system from its correlation with a k-reservoir.
\newblock {\em Journal of Mathematical Physics}, 19(5):1028--1031, 1978.

\bibitem{muller2012unifying}
Markus~P M{\"u}ller, Oscar~CO Dahlsten, and Vlatko Vedral.
\newblock Unifying typical entanglement and coin tossing: on randomization in probabilistic theories.
\newblock {\em Communications in Mathematical Physics}, 316:441--487, 2012.

\bibitem{oliveira2007generic}
R~Oliveira, OCO Dahlsten, and MB~Plenio.
\newblock Generic entanglement can be generated efficiently.
\newblock {\em {Physical Review Letters}}, 98(13):130502, 2007.

\bibitem{page1993average}
Don~N Page.
\newblock Average entropy of a subsystem.
\newblock {\em {Physical Review Letters}}, 71(9):1291, 1993.

\bibitem{page1993information}
Don~N Page.
\newblock Information in black hole radiation.
\newblock {\em {Physical Review Letters}}, 71(23):3743, 1993.

\bibitem{peres1997quantum}
Asher Peres.
\newblock {\em Quantum theory: concepts and methods}, volume~72.
\newblock Springer, 1997.

\bibitem{popescu2006entanglement}
Sandu Popescu, Anthony~J Short, and Andreas Winter.
\newblock Entanglement and the foundations of statistical mechanics.
\newblock {\em Nature Physics}, 2(11):754--758, 2006.

\bibitem{vedral1997quantifying}
Vlatko Vedral, Martin~B Plenio, Michael~A Rippin, and Peter~L Knight.
\newblock Quantifying entanglement.
\newblock {\em Physical Review Letters}, 78(12):2275, 1997.

\end{thebibliography}

\end{document}